\documentclass[epj]{svjour}
\usepackage{amsmath}
\usepackage{amssymb,mathrsfs}
\usepackage{graphicx}% Include figure files

\begin{document}

\title{Interference effects in the two-dimensional scattering of
  microcavity polaritons by an obstacle: phase dislocations and
  resonances}
\titlerunning{Interference effects in the two-dimensional scattering of
polaritons}
\author{A. M. Kamchatnov\inst{1} \and N. Pavloff\inst{2}}
\institute{Institute of Spectroscopy,
  Russian Academy of Sciences, Troitsk, Moscow Region, 142190, Russia \and
Univ. Paris Sud, CNRS, Laboratoire de
  Physique Th\'eorique et Mod\`eles Statistiques, UMR8626, F-91405
  Orsay, France}

\abstract{
  We consider interference effects within the linear description of
  the scattering of two-dimensional microcavity polaritons by an
  obstacle. The polariton wave may exhibit phase dislocations created
  by the interference of the incident and the scattered fields. We
  describe these structures within the general framework of singular
  optics. We also discuss another type of interference effects
  appearing due to the formation of (quasi)resonances in the potential
  of a repulsive obstacle with sharp boundaries. We discuss the
  relevance of our approach for the description of recent experimental
  results and propose a criterion for evaluating the importance of
  nonlinear effects.
}
\maketitle

\section{Introduction}

In a recent publication \cite{cilibrizzi} Cilibrizzi {\it et al.}
reported experimental and theoretical results on the scattering of a
two-dimensional (2D) flow of microcavity polaritons by a localized
potential.  A specific wave pattern was identified in the wake of this
obstacle: elongated regions of low density would separate brighter
zones, and the phase of the wave function would experience rapid jumps
across the low density regions.  As was indicated in
Ref. \cite{cilibrizzi}, these features are reminiscent of the
nonlinear oblique solitons generated by the two-dimensional supersonic
flow of a Bose-Einstein condensate past an obstacle. Such nonlinear
structures were predicted and analyzed in Refs.~\cite{egk-2006}
and \cite{Kam08} for atomic condensates, and observed
experimentally in the flow of microcavity polariton past an obstacle
\cite{amo-2011,grosso-2011}. As the structures observed in
Ref.~\cite{cilibrizzi}, oblique solitons manifest themselves as
strips of diminished density connecting regions with markedly
different phases. However, the range of density in the experiment of
Ref.~\cite{cilibrizzi} was such that nonlinear effects could
safely be discarded in its theoretical modeling. This led Cilibrizzi
{\it et al.} to question the nonlinear paradigm used in the
interpretation of Refs.~\cite{amo-2011} and
\cite{grosso-2011}, a claim which has been itself objected in
the Comment \cite{comment}.

Ignoring for a moment the controversy on the observation of oblique
solitons in Refs.~\cite{amo-2011} and \cite{grosso-2011}, it remains
that Ref.~\cite{cilibrizzi} displays interesting results which deserve
a clear interpretation. Ideally this interpretation should suggest
experimental signatures making it possible (i) to discriminate the
linear and the nonlinear regimes and (ii) to observed new effects in
the field of microcavity polaritons. This is the goal of the present
work: we model the 2D polaritonic flow without account of nonlinear
effects and attribute---as Cilibrizzi {\it et. al.} already did---the
specific features observed in Ref.~\cite{cilibrizzi} to phase
singularities which mimic some of the aspects of oblique solitons. We
also propose criteria allowing to attribute specific characteristics
to oblique solitons (not seen in the linear case) and some others to
linear interference effects. We furthermore propose to extend the
range of parameters of the linear experiment in order to demonstrate
some peculiar effects of linear 2D scattering in the new framework of
microcavity polaritons.

The paper is organized as follows: in Sec.~\ref{section.model} we
present the model we use in the paper and draw first a qualitative
then a quantitative picture of the scattering process. In
Sec.~\ref{section.dislocation} we present the simpler 2D phase
singularity: the edge dislocation and identify such structures in the
wake of a 2D obstacle. In Sec.~\ref{section.resonant} we discuss an
other linear wave effect connected to quasi-resonant
scattering. Finally we present our conclusions in
section~\ref{section.conclusion}.

\section{The linear wave model}\label{section.model}

In the 2D geometry
appropriate for the description of
a planar microcavity, the linear dynamics of the polariton field
$\psi(\vec{r},t)$ can be simply described by the Schr\"odinger equation
\begin{equation}\label{eq1}
    {\rm i}\,\hbar\, \psi_t=
-\frac{\hbar^2}{2m}\nabla^2\psi
+U(\vec{r})\,\psi,
\end{equation}
where $\vec{r}=(x,y)=(r,\varphi)$ locates the position in the plane
and $m$ is the effective mass of polaritons. $U(\vec{r})$ is the
scattering potential which we assume to be of finite extend and
localized near the origin.  For simplicity, we have neglected here
all dissipation and pumping effects. We aim at studying a
configuration where polaritons are injected ahead the obstacle and
propagate with a single wave-vector $k$. We typically consider the
case where the incident beam is a plane wave, but we also present
results pertaining to the case of a circular wave emitted by a point
source (the incident plane wave corresponds to the limiting case where
the source is at infinity).  In these configurations, the solution of
Eq.~\eqref{eq1} is a stationary function $\psi(\vec{r})
\exp(-{\rm i} E t/\hbar)$ (with $E=\hbar^2 k^2/2m$) which can be
decomposed into incident and scattered parts.

As stated in the introduction, we suppose that the structures observed
in Ref.~\cite{cilibrizzi} can be interpreted as manifestations
of singular optics effects \cite{nye99,sv-2001}, namely, the
appearance of {\it phase dislocations} \cite{nb-1974} in the wave
pattern produced by the interference of the incident and scattered
waves.  Such structures have been observed in various physical
contexts (see, e.g., Refs.~\cite{nye99} and \cite{sv-2001}
and references therein) and they result in elongated dips in the
density distributions and sharp changes of the phase in vicinity of
amplitude nodal points. Highly anisotropic density dips appear for
instance around nodal points in the interference pattern issued form
the diffraction of a plane wave from a reflecting half line, see e.g.,
Fig. 11.12 of Ref. \cite{BornWolf} or
Figs. 1 and 2 of Ref. \cite{Berry330}.

Since the low density region around a node in the wake behind the
obstacle can have a very elongated form, it can look like part of an
oblique soliton's strip. To qualitatively illustrate this idea, let us
first consider the situation of $s$-scattering by an obstacle
described by a constant (i.e., $\varphi$ independent) amplitude $f$.
Far enough from the obstacle the polariton field can be represented in
the form \cite{LL-3}
\begin{equation}\label{eq2}
    \psi(\vec{r})=e^{{\rm i}kx}+\frac{f}{\sqrt{r}} \, e^{{\rm i}(kr+\pi/4)},
\end{equation}
where the incident wave propagates along the positive $x$ axis. If
$f^2\gg2\pi/k$ and if the range $a$ of the obstacle's potential is
small enough ($a\ll f^2$), then the asymptotic formula (\ref{eq2}) is
valid already for $r\lesssim f^2$. As one can easily see, in this case
Eq.~(\ref{eq2}) yields lines of constant phase with different
topologies: (i) for $r\ll f^2$ the second term in the right hand side
of (\ref{eq2}) dominates and these lines are closed curves (nearly
circles), whereas (ii) they become open lines approaching horizontal
lines for $r\gg f^2$. Hence in the transient region $r\sim f^2$ the
lines of constant phase (the wave fronts) must change topology. This
can be realized through the occurrence of point-like phase
singularities whose specific properties are similar to those observed
in Ref.~\cite{cilibrizzi}.

To study these effects quantitatively, we shall
consider an obstacle represented by a 2D
circular square well potential
\begin{equation}\label{eq3}
    U(\vec{r})=\left\{
          \begin{array}{ccc}
        U_0 & \mbox{for} & r<a \, ,\\
        0 & \mbox{for} & r>a \, , \\
      \end{array}
      \right.
\end{equation}
where $U_0$ can be either positive (for an attractive potential)
or negative (for a repulsive one). We shall start with the simple situation
of a weak potential for which perturbation theory can be applied.

\subsection{Born approximation}

In the most general case, the scattering amplitude $f$ in
Eq. \eqref{eq2} is a function of the angle $\varphi$ between the
incident wave vector $\vec{k}$ (which we choose directed along the
positive $x$ axis) and the wave vector $\vec{k}'$ of the scattered
field. If the potential satisfies the condition
\begin{equation}\label{eq4}
    |U_0|\ll\frac{\hbar^2}{ma^2},
\end{equation}
then we can evaluate the amplitude $f(\varphi)$ within the Born
approximation \cite{LL-3}:
\begin{equation}\label{eq5}
    f(\varphi)=-\frac{m}{\hbar^2\sqrt{2\pi k}}\int
    U(\vec{r})e^{-i\vec{q}\cdot \vec{r}}{\rm d}^2\vec{r},
\end{equation}
where $\vec{q}=\vec{k}'-\vec{k}$, which yields in our case of
elastic scattering
\begin{equation}\label{eq6}
    q=|\vec{q}|=2k\sin(\varphi/2).
\end{equation}
Applying equation \eqref{eq5} to the potential \eqref{eq3} and using
the well-known formulae
\begin{equation}\label{eq7}
    J_0(z)=\frac1{2\pi}\int_0^{2\pi}\!\! e^{-{\rm i}z\cos\theta}{\rm d}\theta,\;
    \int_0^a\!\! z J_0(z){\rm d}z=a J_1(a)
\end{equation}
for the Bessel functions $J_n(z)$, we obtain
\begin{equation}\label{eq8}
    f(\varphi)=-\frac{m a^2 U_0}{\hbar^2}\sqrt{\frac{2\pi}k}\cdot
    \frac{J_1(2 a k\sin(\varphi/2))}{2 a k\sin(\varphi/2)}.
\end{equation}
The oscillatory behavior of the function $J_1(2a k\sin(\varphi/2))$
leads to the appearance of ``valleys'' of diminished density in the
interference pattern corresponding to (\ref{eq2}), a feature which
agrees qualitatively with the wave patterns observed in
Ref.~\cite{cilibrizzi}. However, the depth of these valleys is small
because of the condition (\ref{eq4}) and for getting a more realistic
description of the phenomenon we have to turn to the exact solution of
the scattering problem under consideration.

\subsection{Exact solution}

Exact solutions describing the scattering of sound and electromagnetic
waves on cylindrical obstacles were obtained long ago by Lord Rayleigh
\cite{rayleigh-1,rayleigh-2} and we shall apply the same method to the
polariton field described by the Schr\"odinger equation (\ref{eq1})
with the potential (\ref{eq3}).  We assume that the potential is
either attractive ($U_0<0$) or repulsive but with a potential energy
smaller than the kinetic energy of incident polaritons:
$0<U_0<\hbar^2k^2/(2m)$ (this limiting assumption is made for
simplifying the presentation, but the method equally applies for
$U_0>\hbar^2k^2/(2m)$, see the end of the present section and
Sec. \ref{section.resonant}). Then the wave vector in the
region occupied by the obstacle ($r<a$) is
\begin{equation}\label{eq9}
k_0=\sqrt{k^2-2mU_0/\hbar^2}\; .
\end{equation}
If we assume that polaritons are emitted by a point-like source
located outside of the radius of the potential at the point with
cylindrical coordinates $\vec{r}_1=(r_1,\varphi_1)$, we are actually
interested
in calculating the Green function $G(\vec{r},\vec{r}_1)$,
[where $\vec{r}=(r,\varphi)$ is the radius vector of the observation
point] of the stationary Schr\"odinger equation. $G$ is solution of
the equation
\begin{equation}\label{eq10}
\left\{
\begin{array}{lcc}
\left(\nabla^2_{\vec{r}} +k^2\right)G=4\,{\rm i}\,
\delta(\vec{r}-\vec{r}_1)
& \mbox{if} & r>a \; ,\\
\left(\nabla^2_{\vec{r}} +k_0^2\right)G=0
& \mbox{if} & r<a \; .
\end{array}\right.
\end{equation}
In this equation we added a factor $4{\rm i}$ in the source
term: this is a simple aesthetic modification
allowed by the linearity of the problem.
In the absence of potential the first of Eqs.~\eqref{eq10} is valid in
whole space and the associated causal Green function is the Hankel
function $H^{(1)}_0 (k|\vec{r}-\vec{r}_1|)$. It is thus appropriate to
look for a solution of \eqref{eq10} of the form
\begin{equation}\label{eq10bis}
G(\vec{r},\vec{r}_1) =H^{(1)}_0 (k|\vec{r}-\vec{r}_1|) \Theta(r-a)+
G_1(\vec{r},\vec{r}_1)\; ,
\end{equation}
where $\Theta$ is the Heaviside function.
$G_1$ is solution of a cylindrical
symmetric problem which can be solved by the method of separation of
variables, yielding the following expression:
\begin{equation}\label{eq12}
 G_1(\vec{r},\vec{r}_1)=
 \left\{
 \begin{array}{l}
 \displaystyle \sum_{n=-\infty}^{\infty}B_nH_n^{(1)}(k r)
 e^{{\rm i}n(\varphi-\varphi_1)},\quad r>a,\\
 \displaystyle \sum_{n=-\infty}^{\infty}A_nJ_n(k_0r)
 e^{{\rm i}n(\varphi-\varphi_1)},\quad r<a\; ,
 \end{array}
 \right.
\end{equation}
where standard notations are used for special functions from the
Bessel family (see, e.g., Ref.~\cite{AS}).
The combination of Bessel functions used in the expression
\eqref{eq12} is chosen in order to satisfy the asymptotic conditions of
the problem: in the region $r>a$, the choice of $H_n^{(1)}(k r)$ ensures
that one considers outgoing waves, and in the region $r<a$, the choice
of $J_n(k_0r)$ ensures that the wave function is not singular at the
origin.

Using the addition formula for Bessel functions \cite{AS}
one may write the Hankel function in \eqref{eq10bis} as (for $r_1>r$)
\begin{equation}
H^{(1)}_0 (k|\vec{r}-\vec{r}_1|) =
\sum_{n=-\infty}^\infty H_n^{(1)}(k r_1)J_n(k r) e^{{\rm i}n(\varphi-\varphi_1)} \; .
\end{equation}
Then, the coefficients $A_n$ and $B_n$ can be found from the conditions of
continuity of the function $G$ and of its derivative at the obstacle
boundary $r=a$. Simple manipulations yield
\begin{equation}\label{eq14}
B_n=\tilde{B}_n \, H_n^{(1)}(k r_1) \; ,\quad
A_n=\tilde{A}_n \, H_n^{(1)}(k r_1) \; ,
\end{equation}
with
\begin{equation}\label{eq20}
\tilde{B}_n=\frac{-k_0J_n^{\prime}(k_0a)J_n(ka)+kJ_n(k_0a)J_n^{\prime}(ka)}
{k_0J_n^{\prime}(k_0a)H_n^{(1)}(ka)-kJ_n(k_0a)H_n^{(1)\prime}(ka)}\; ,
\end{equation}
\begin{equation}\label{eq16}
\tilde{A}_n={\rm i}\frac{kJ_n^{\prime}(ka)Y_n(ka)-kJ_n(ka)Y_n^{\prime}(ka)}
{k_0J_n^{\prime}(k_0a)H_n^{(1)}(ka)-kJ_n(k_0a)H_n^{(1)\prime}(ka)} \; ,
\end{equation}
where the prime denotes the derivative functions.

If the source is located at $x_1\to-\infty$ and the incoming wave is
represented by the plane wave function $e^{{\rm i}kx}$, then the
formulae can be simplified.  The Green function takes the form
\footnote{Note that in the presence of losses the source term could
  have a wave vector different from $k$. In this case, different types
  of wave pattern can be observed, as discussed by \cite{remarkiac}.}
\begin{equation}\label{eq17a}
G(\vec{r},\vec{k})=\exp({\rm i} k x)  \Theta(r-a)
+ G_1(\vec{r},\vec{k})\; ,
\end{equation}
with
\begin{equation}\label{eq17}
 G_1(\vec{r},\vec{k})=
 \left\{
 \begin{array}{l}
 \displaystyle \sum_{n=-\infty}^{\infty} {\rm i}^n \tilde{B}_n
 H_n^{(1)}(k r)\, e^{{\rm i}n \varphi}\quad r>a\; ,\\
 \displaystyle \sum_{n=-\infty}^{\infty} {\rm i}^n\tilde{A}_n
 J_n(k_0r)\, e^{{\rm i}n \varphi}\quad r<a\, .
 \end{array}
 \right.
\end{equation}
In the case of a repulsive obstacle with $U_0>\hbar^2 k^2 /(2 m)$, the
above treatment still holds but $k_0$ should now be defined as
\begin{equation}\label{eq9bis}
k_0={\rm i} \sqrt{-k^2+2\,m\, U_0/\hbar^2}\; .
\end{equation}
If one considers a hard disk scatterer, i.e., an infinitely repulsive
obstacle of radius $a$, then $k_0$ in \eqref{eq9bis}
tends to ${\rm i}\infty$, but the
expressions \eqref{eq12}, \eqref{eq14} and \eqref{eq17} remain valid
provided \eqref{eq20} and \eqref{eq16} are replaced by
\begin{equation}\label{eq13}
\tilde{B}_n=-\frac{J_n(ka)}{H_n^{(1)}(ka)} \; , \quad \tilde{A}_n=0 \; .
\end{equation}

The formulae \eqref{eq12} to \eqref{eq13} give the exact solution of
our scattering problem, which, in the limit $k r\to\infty$ and for an
incident plane wave, takes the asymptotic form (\ref{eq2}), where the
scattering amplitude is here given by the expression
\begin{equation}\label{eq19}
f(\varphi)=-{\rm i}\, \sqrt{\frac2{\pi k}}\sum_{n=-\infty}^{\infty}
\tilde{B}_n\, e^{{\rm i}n\varphi} \; .
\end{equation}
If the condition (\ref{eq4}) is fulfilled, then after tedious
manipulations with the use of Graf's addition theorem and recurrence
relations for Bessel functions one can reproduce the result
(\ref{eq8}) of the Born approximation starting from the expressions
\eqref{eq19} and \eqref{eq20}. The complexity of these manipulations
illustrates the fact that the partial wave expansion \eqref{eq12} and
\eqref{eq17} is not adapted to a perturbative approach. It is however
very well suited for a numerical treatment of the problem: in
practice, the partial wave expansion can be limited to a range $|n|\le
{\cal O}(k\, a)$ which makes the numerical determination of the wave
function fast and easy.

\begin{figure}
\includegraphics*[width=\linewidth]{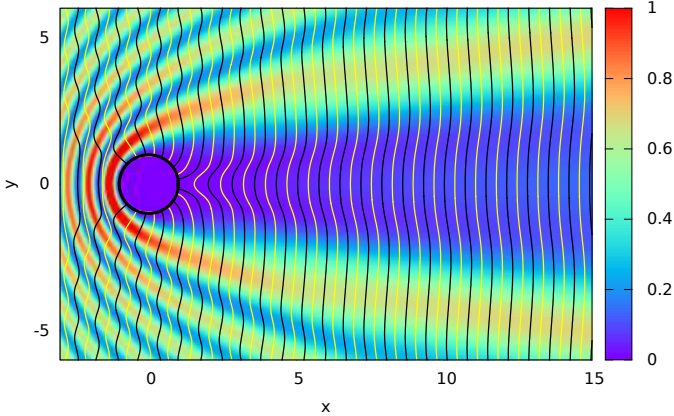}
\caption{Color plot of the density (in arbitrary units)
  corresponding to the Green function \eqref{eq17} in the case of a
  hard disk scatterer of radius $a=1$. The incident wake vector is
  $\vec{k}=4.5\,a^{-1} \vec{e_x}$, corresponding to the
  situation studied in Ref.~\cite{cilibrizzi} ($ka=4.5$). The
  yellow solid lines are the lines of equiphases $0$ and $\pi$ and the
  black solid lines are the lines of equiphase $\pm
  \pi/2$.}\label{fig1}
\end{figure}

\begin{figure}
\includegraphics*[width=\linewidth]{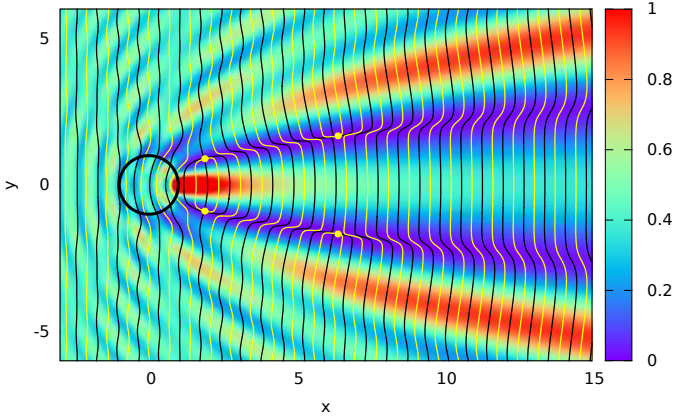}
\caption{Same as Fig.~\ref{fig1} for an attractive
  potential with $2ma^2U_0/\hbar^2=-15$. The four yellow points are
  zeros of the wave function. They are approximately located at
  positions $(1.899,\pm 0.8928)$ and $(6.4056,\pm
  1.674)$.}\label{fig2}
\end{figure}

As an illustration of the numerical method, we display in
Figs.~\ref{fig1} and \ref{fig2} a color plot of the intensity of the
Green function (\ref{eq17a},\ref{eq17}) in the case $k\,
a=4.5$. Fig.~\ref{fig1} corresponds to a hard disk scatterer and
Fig.~\ref{fig2} to a penetrable attractive potential with
$2ma^2U_0/\hbar^2=-15$. The figures are plotted restraining the
summations in \eqref{eq12} and \eqref{eq17} to $|n|\le 10$. We checked
that including higher partial waves does not modify the figure. We
also display wavefronts in the figures: the yellow solid lines are the
lines of equiphases $0$ and $\pi$ (i.e., the zeros of
$\mathrm{Im}\,G$) and the black solid lines are the lines of equiphase
$\pm \pi/2$ (i.e., the zeros of $\mathrm{Re}\,G$).

The diffraction pattern of Fig.~\ref{fig1} displays no noticeable
structure: the wake of the (impenetrable) potential corresponds to the
shadow of the obstacle and to a region of low density. On the other
hand, in the case of a penetrable (attractive) potential, one sees a
region of high density (for $x>0$ and $y\simeq 0$) separating two
elongated regions of low density (cf. Fig.~\ref{fig2}). These low
density regions are similar to the ones observed in
Ref.~\cite{cilibrizzi} and are located around zeros of the wave
function. The zeros in Fig. \ref{fig2} are indicated by four yellow
points at which the yellow and black wavefront (respectively zeros of
$\mathrm{Im}\,G$ and of $\mathrm{Re}\,G$) cross. They are associated
to phase singularities, as we now discuss.

\section{Wave singularities}\label{section.dislocation}

While the scattering pattern of Fig.~\ref{fig1} has no noticeable
structure, one sees zeros of the wave function in the wake of the
obstacle of Fig.~\ref{fig2}. They are easily located as points where
the lines $\mathrm{Re}\,G=0$ and $\mathrm{Im}\,G=0$ cross. In a first
part of this section we present the local behavior of $\psi$ around the
nodal points, and in the second part we verify that the exact solution
\eqref{eq17} has all the characteristic behaviors identified in
Sec.~\ref{sec.model}.

\subsection{Model Case}\label{sec.model}

\begin{figure}
\includegraphics*[width=0.45\linewidth]{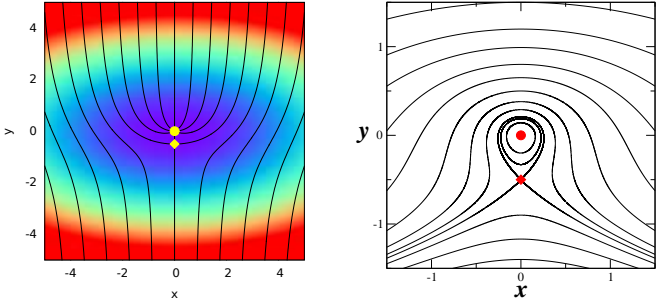}
\hspace{0.05\linewidth}
\includegraphics*[width=0.45\linewidth]{fig3b}
\caption{Left plot: color plot of the density
  corresponding to the wave function \eqref{eq21} ($\alpha=0.5$ and
  $k=1$). The blue (red) color corresponds to a region of lower
  (higher) density. The black solid lines are wavefronts
  ($\theta=0,\pm \pi/4,\pm \pi/2,\pm 3\pi/4\pi,\pi)$.  The nodal point
  at the origin is marked by a yellow circle and the stagnation point
  \eqref{eq25} by a yellow diamond. Right plot: corresponding
  streamlines. Note that the scale (and the color of the singular
  points) has been modified in order to get a better resolution in the
  region near the origin.}\label{fig3}
\end{figure}

Let us denote as $X$ and $Y$ the abscissa and ordinate
in a coordinate system whose origin is fixed at a nodal point of the
wave function. We chose the $X$-axis in such a way that the
wave field can be locally represented in the form
\begin{equation}\label{eq21}
 \psi\cong (\alpha X-{\rm i}Y)\, e^{{\rm i}k X},
\end{equation}
that is, the wave is locally represented by a plane wave whose
amplitude vanishes at the origin. The coefficient $\alpha>0$ controls
the scales along the coordinate axes and the choice of signs is made
for later convenience.  As one can easily see, the density
distribution
\begin{equation}\label{eq22}
 |\psi|^2=\alpha^2 X^{2}+Y^{2}
\end{equation}
has an elliptic form, and if $\alpha\ll1$ the lines of constant
density are strongly elongated along the $X$-axis. The phase has a
more interesting behavior corresponding to an {\it edge dislocation}
in the wave field \cite{nb-1974}.  The phase $\theta$ of the wave
function is defined by the equation
\begin{equation}\label{eq23}
 \theta(X,Y)=\arctan\frac{\alpha X\sin(kX)
 -Y\cos(kX)}{\alpha X\cos(kX)+Y\sin(kX)},
\end{equation}
We show in Fig.~\ref{fig3} wavefronts (i.e., lines of constant phase) and
streamlines which are the field lines of the vector field $\nabla\theta$
with
\begin{equation}\label{eq24}
\theta_X=
\frac{\alpha(Y+\alpha kX^2)+k Y^2}
{|\psi|^2},\quad
\theta_Y=-\frac{\alpha X}{|\psi|^2}.
\end{equation}
One can see that the lines of constant phase stem from the origin and
that the streamlines form circles around it. This means that there is
a vortex located at the origin (the nodal point) and that the velocity
field $\nabla\theta$ has a singularity here. There is another
remarkable point where the backward velocity induced by the vortex
just cancels the plane wave flow velocity $k\, \vec{e_x}$. This is
a critical point of the vector field (\ref{eq24}) with vanishing
velocity $(\theta_X,\theta_Y)$.  Its coordinates are given by
\begin{equation}\label{eq25}
    X=0\; ,\quad Y=-\alpha/k\; .
\end{equation}
It is easy to find that the Hessian has here opposite eigenvalues
($\pm\alpha^3/k^2$), meaning that this critical point is a saddle, as
clearly seen in the left plot of Fig.~\ref{fig3}.  The separatrix
going through the saddle point has a fixed phase which can be put
equal to $\pi$, or equal to $\pi$ along one branch and equal to
$-\pi\equiv\pi\mod(2\pi)$ along another branch. Therefore the phases
of the wavefront at this point can change values between two choices equal
to each other modulus $2\pi$ and the whole change of phase depends on
the charge of the vortex located ``inside'' the separatrix line.  In
our case (Fig. \ref{fig3}) the phase changes from $-\pi$ at one branch of the
separatrix to $\pi$ at the other branch (if one goes from left to
right), that is the charge of the vortex is equal to unity.

\begin{figure}
\includegraphics*[width=0.45\linewidth]{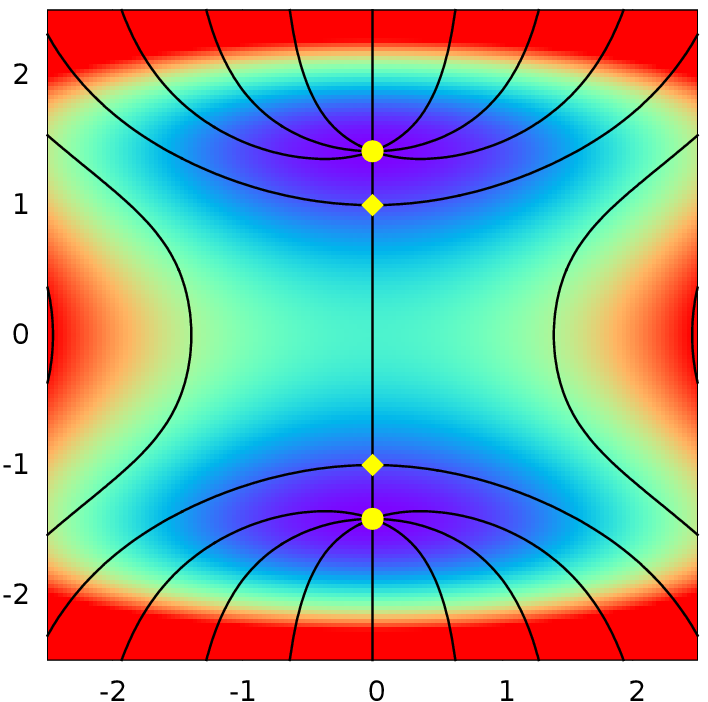}
%\hspace{0.01\linewidth}
\hfill
\includegraphics*[width=0.45\linewidth]{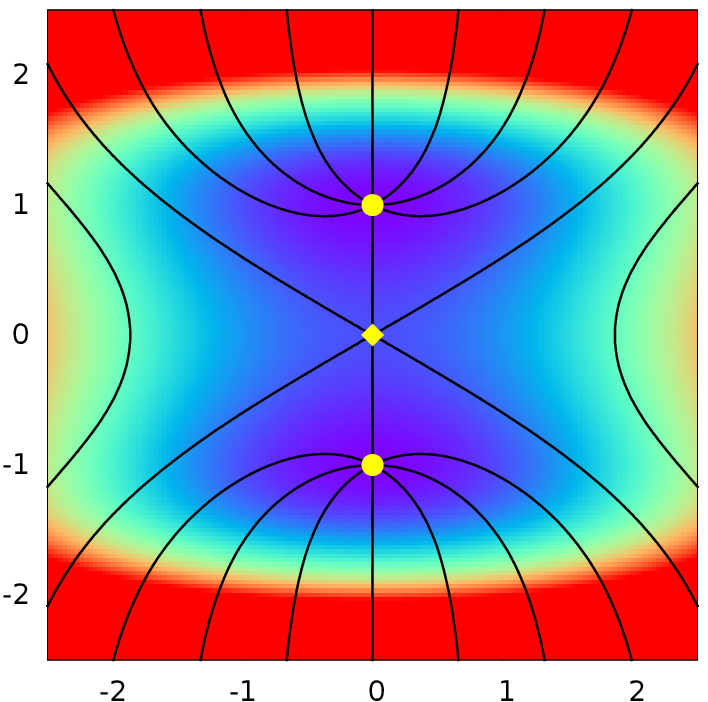}
\includegraphics*[width=0.45\linewidth]{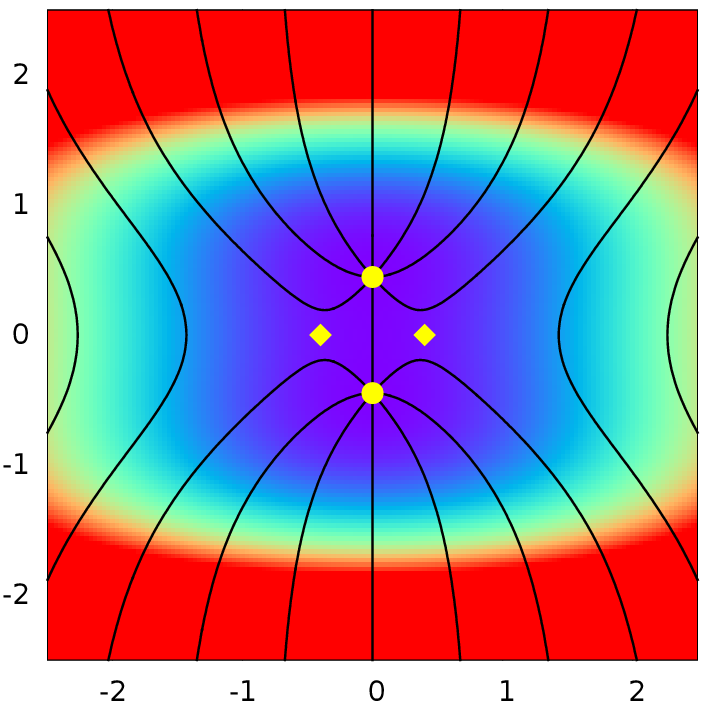}\hfill
%\hspace{0.01\linewidth}
\includegraphics*[width=0.45\linewidth]{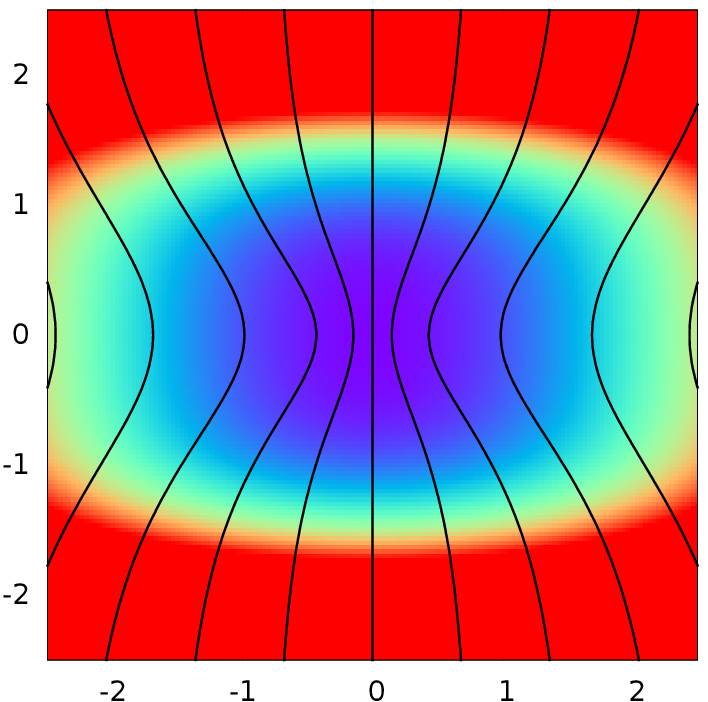}
\caption{Plot of the intensity of the wave
  function \eqref{eq26} (with $k=1$) for different values of the parameter $b$
  (from upper left to lower right: $b=2$, 1, $0.2$ and $-0.2$). The blue (red)
  color corresponds to a region of lower (higher) density. In each plot
the black solid lines are the wavefronts
  $\theta=0,\pm \pi/4,\pm \pi/2,\pm 3\pi/4\pi,\pi$, the yellow circles
  are the nodal point and the diamonds are the saddle points.}\label{fig4}
\end{figure}

As one can see, the edge phase singularity described here leads to a
very elongated region of low density if $\alpha\ll1$
[cf. (\ref{eq22})] which can mimic the experimental situation observed
in Ref.~\cite{cilibrizzi} if the distance $k/\alpha$ is less
than the typical distance between the ``pseudo-soliton'' and the
incident flow axis.
If this condition is not fulfilled, then we have to take
into account the existence of a symmetrical edge phase singularity below
the axis of the flow. This can be done by making the approximation that the
angles between the ``pseudo-solitons'' and the axis of the flow are negligibly
small. In this case we can approximate locally the solution of our
scattering problem by the exact solution of the Helmholtz equation
(see Refs.~\cite{nb-1974,nye99,Nye88})
\begin{equation}\label{eq26}
    \psi=\left[X-{\rm i}k(Y^2-b)\right]e^{{\rm i}kX}\; ,
\end{equation}
where $b$ is real. If $b$ is positive, this field has nodes at points
$(0,\pm \sqrt{b})$ corresponding to vortices with opposite circulations.
The corresponding velocity field $\nabla\theta$ has two stagnation
points with coordinates $(X_s,Y_s)$
\begin{equation}\label{eq27}
    (X_s,Y_s)=\left\{
    \begin{array}{lcc}
    (0,\pm \sqrt{b-k^{-2}}) & \text{if} & b>k^{-2},\\
    (\pm \sqrt{b-k^2b^2},0) & \text{if} & b<k^{-2},
    \end{array}
    \right.
\end{equation}
and it is easy to show that these are saddle points. Thus, for large
enough values of $b$ the two vortices and the two saddle points are
located on the $y$ axis and the resulting structure can be represented
as a symmetrical combination of two edge phase dislocations with a
shallow low density region of large horizontal extension (see the
upper left plot of Fig. \ref{fig4}). When $b$ decreases, an interesting
mechanism of collapse of the singularities occurs which is depicted in
Fig.~\ref{fig4}. For $b=k^{-2}$ (upper right plot) the saddle points
collide at the origin and for decreasing $b$ they start to move
symmetrically from the origin along the $x$ axis (lower left plot):
after initially drifting apart, the two saddles get closer anew. At
last, when $b=0$ all vortices and saddle points annihilate at the
origin, the dislocation disappears and for $b<0$ the flow becomes
regular (lower right plot).

This scenario of disappearance of the wave singularities was put
forward by Nye, Hajnal and Hannay in Ref.~\cite{Nye88}. It is
robust because it obeys the restrictions dictated by topology, as we
now explain. Two topological indices can be ascribed to the nodal and
saddle points. One of them, known as the topological charge or
vorticity $I_{\rm V}$, measures the circulation of the velocity field around the
singularity: $I_{\rm V}=\pm 1$ around the nodes of the wave functions
\eqref{eq21} and \eqref{eq26} and $I_{\rm V}=0$ around a saddle. The other
one, known as Poincar\'e index $I_{\rm P}$, measures the change of direction of
the equiphase lines around the singularity: $I_{\rm P}=1$ for a node
and $I_{\rm P}=-1$ for a saddle. The simple annihilation of two
vortices of opposite topological charge is not possible because it
does not conserve the Poincar\'e index. In the Nye {\it et al.} scenario
instead, the concomitant annihilation of the zeros and of the saddles
conserve both the total vorticity and Poincar\'e index.

\subsection{The wake of a penetrable disk}

\begin{figure}
\includegraphics*[width=0.99\linewidth]{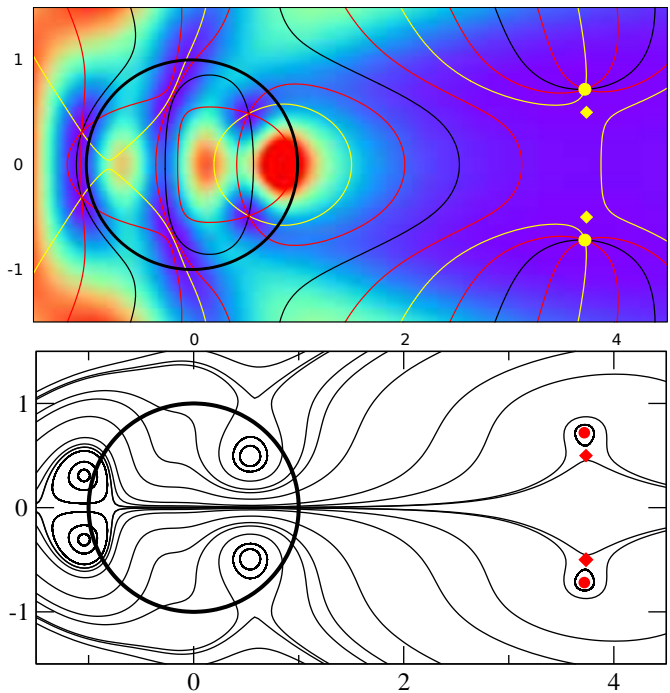}
\hspace{0.01\linewidth}
\includegraphics*[width=0.99\linewidth]{fig5b}
\caption{Upper plot: density of the wave function for a beam
  scattering onto a penetrable disk with $2 m a^2 U_0/\hbar^2=-15$. The
  horizontal (vertical) axis is the $x$-axis ($y$-axis) in units of
  $a$. The color code is the same as in Figs. \ref{fig1} and
  \ref{fig2}. The incident wave vector is $\vec{k}= \frac{2.0}{a}\,
  \vec{e_x}$. The yellow solid lines are the lines of equiphases
  $0$ and $\pi$, the black solid lines are the lines of equiphase $\pm
  \pi/2$ and the purple solid lines are the lines of equiphase $\pm
  \pi/4$ and $\pm 3\pi/4$. Lower plot: streamlines. The nodes and
  saddles in the region $x\simeq 3.7$ are marked by yellow (red)
  points and diamonds in the upper (lower) plot. The positions of the
  other nodes and saddles are not marked by a special signs in the
  figure.}\label{fig5}
\end{figure}

\begin{figure}
\includegraphics*[width=0.99\linewidth]{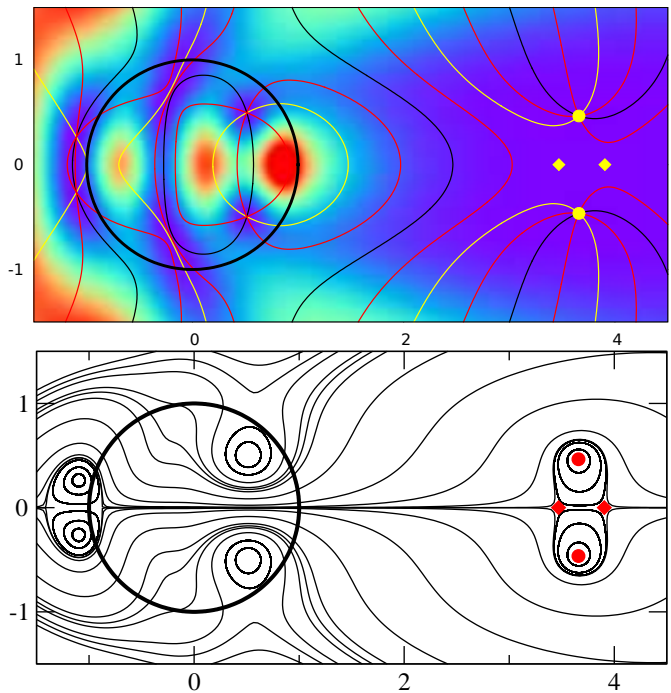}
\hspace{0.01\linewidth}
\includegraphics*[width=0.99\linewidth]{fig6b}
\caption{Same as Fig.~\ref{fig5} for $k=1.9/a$.}\label{fig6}
\end{figure}

In our scattering problem the process just described is controlled by
the parameters $U_0$ (the depth of the potential) and $k$ (the
incident wave vector). We now show that this process is indeed
observed when changing the parameters $U_0$ and $k$ starting from the
situation depicted in Fig.~\ref{fig2}. In view of a possible
experimental implementation it is more appropriate to keep $U_0$ fixed
and to change the value of the incident wave vector. Fig.~\ref{fig5}
depicts the density pattern and the corresponding streamlines for
$k=2.0/a$ (and $2 m a^2 U_0/\hbar^2=-15$).
For this value of $k$ the zeros of the wave function in the wake of
the obstacle are closer one to each other than in Fig. \ref{fig2}. One
is in a situation similar to that shown in the upper left plot of
Fig.~\ref{fig4}: Two nodes and two saddles are almost aligned on a
vertical line\cite{remark_saddles} in the region $x\simeq 3.7\, a$. If
$k$ slightly decreases to $k=1.9/a$, one obtain the results depicted
in Fig.~\ref{fig6}: the two nodes still have symmetric positions with
respect to the horizontal axis but the saddles now lay on this
axis. This is a situation similar to the one depicted in the left
lower plot of Fig. \ref{fig4}. Finally, for $k=1.8/a$ (not shown) the
flow becomes regular in the region $x\simeq 3.7\,a$ and $y=0$: Hence
the disappearance of the zeros and of the saddles in this region
exactly follows the scenario of Nye, Hajnal and Hannay presented in
the previous sub-section.

The detailed analysis just presented of the symmetrical regions of
lower density in the wake of an attractive obstacle (depicted in
Fig.~\ref{fig2}) confirms the original discussion of
Ref.~\cite{cilibrizzi}: these zones correspond to phase
singularities in vicinity of zeros of the wave function. Indeed, the
rapid change of the phase across these regions mimics the behavior of
the phase across the oblique solitons observed in the wake of a
nonlinear fluid.

However, there are marked differences between the linear phase
singularities and the oblique solitons: (a) the phase singularities
are not seen in the wake of an impenetrable obstacle (see
Fig.~\ref{fig1}); (b) the phase singularities are less robust, in the
sense that they can easily disappear upon changing the incident wave
vector or the depth of the potential, however, this disappearance is
in itself an interesting phenomenon governed by topological
constraints as just verified; (c) the phase singularities occupy only
a finite region of the stationary interference pattern whereas the length
of oblique solitons increases with time in principle {\it ad
  infinitum} (see Ref. \cite{kk-2011}); (d) the width of an oblique
soliton is controlled by the balance of the dispersion and
nonlinearity whereas the characteristic size of the dark
regions around the phase singularities depends on the parameters of
the obstacle only (see the precise discussion in Section
\ref{section.conclusion}); (e) at last, the threshold wave number of
the incident wave for appearance of oblique solitons is related with
the characteristic {\it sound velocity} in the polariton condensate
\cite{Kam08,kk-2011} whereas there is no such threshold for the
formation of a linear interference pattern.

\section{Resonant scattering}\label{section.resonant}

So far we have concentrated our attention on the wave pattern outside
the obstacle where phase dislocations can be formed which are
accompanied by elongated dips in the density distributions. However,
the exact solution (\ref{eq19}), (\ref{eq20}) provides also valuable
informations about the wave distribution {\it inside} the region
occupied by the obstacle.

\subsection{Low energy scattering}

A first interesting situation occurs for low energy scattering on an
attractive potential ($U_0<0$). One works in the low incident energy
regime where $k a\ll 1$ and $\hbar^2k^2/2m \ll |U_0|$. In this case
one can show that the coefficients $\tilde{B}_n$ of the partial wave
expansion in \eqref{eq20} behave as $(ak)^{2n}$ and only the $s$-wave
contributes significantly to the scattering.

It is convenient to measure the depth of the potential well in units
of a wave vector $Q_0$ defined by
\begin{equation}\label{les1}
\frac{\hbar^2 Q_0^2}{2\, m}=|U_0| \; .
\end{equation}
In this case, defining the quantity $\kappa$ by
\begin{equation}\label{les2}
\ln\left( \frac{\kappa\,  a \, e^\gamma}{2}\right) =
\frac{1}{Q_0 a} \, \frac{J_0(Q_0 a)}{J_0'(Q_0 a)} \; ,
\end{equation}
where $\gamma$ is the Euler-Mascheroni constant, and using the
asymptotic expansions of Hankel and Bessel functions for small
argument \cite{AS} one obtains from \eqref{eq20}:
\begin{equation}\label{les3}
\tilde{B}_0\simeq \frac{-1}{1+\frac{2{\rm i}}{\pi}\ln(k/\kappa)} \; .
\end{equation}
A similar expression is generally valid for any low energy scattering
process in two dimensions\cite{LL-3}. Eq. \eqref{les3} also applies
for a repulsive potential, but in this case the definition
\eqref{les2} of $\kappa$ should be replaced by \eqref{cross2}. In the
case of a hard disk scatterer for instance, the constant $\kappa$
takes the value $\kappa=2 e^{-\gamma}/a$: this can be obtained by
taking the limit $U_0\to +\infty$ in \eqref{cross2}, or directly from
the expression \eqref{eq13}. In the attractive case which we consider
in the present sub-section, $\kappa$ has the following physical
meaning: if there is a bound $s$-state close the
threshold\cite{remark6}, i.e., if one is in the case of quasi-resonant
scattering, then this state has an energy $-\hbar^2 \kappa^2/(2m)$.

From \eqref{les3} one sees that the cross section
\begin{equation}\label{les4}
  \sigma=\int_0^{2\pi}|f(\varphi)|^2 {\rm d}\varphi
\stackrel{ka\to 0}{\simeq} \frac{4}{k} \, |\tilde{B}_0|^2 \; ,
\end{equation}
diverges at low energy. Thus, in this limit, the linear scattering
process is markedly different from the nonlinear one for which
superfluidity prevails at low incident velocity. Actually this is
again the manifestation of the existence of an important
characteristic quantity of the polariton condensate, namely the {\it
  velocity of sound}: scattering disappears for a nonlinear flow whose
velocity is lower than the critical velocity which -- in weakly
interacting polariton gas -- is of the order of the sound velocity
\cite{pol-super}.

\subsection{Quasi-stationary states over a repulsive potential}

The scattering amplitude (\ref{eq20}) has poles for complex values of
the variable $k$ and this means that quasistationary states can be
formed under certain conditions. It is interesting to consider such a
possibility here because these states can be detected experimentally.

The poles correspond to zeroes of the denominator in the expression
(\ref{eq20}) of the scattering amplitude, that is they are determined
by the equation
\begin{equation}\label{eq28}
    k_0J_n^{\prime}(k_0a)H_n^{(1)}(ka)=kJ_n(k_0a)H_n^{(1)\prime}(ka).
\end{equation}
This equation has, generally speaking, complex roots with comparable
real and imaginary parts. However, for observing a long living
quasistationary state the imaginary part must be much smaller than the
real one. Such a configuration appears if the repulsive potential
(\ref{eq3}) is large enough,
\begin{equation}\label{eq29}
    U_0\gg \frac{\hbar^2}{ma^2},
\end{equation}
and if the kinetic energy of incident polaritons is only slightly greater
than $U_0$:
\begin{equation}\label{eq30}
    \frac{\hbar^2k^2}{2 m}- U_0\ll U_0 \; .
\end{equation}
Then  we see from Eq.~(\ref{eq9}) that in this case $k_0\ll k$ and the
left hand side of Eq.~(\ref{eq28}) is small, hence $J_n(k_0a)$ must be
accordingly small. This occurs if $k_0a$ is close to a zero of the
Bessel function $J_n(z)$. If, for instance, $k_0a$ is close to the
value $j_{n,s}$ of the $s^{\rm th}$ zero of $J_n$,
\begin{equation}\label{eq31}
    k_{0|n,s}a=j_{n,s}+\delta_{n,s}
\quad (\,\mbox{with}\; |\delta_{n,s}|\ll 1\,) \; ,
\end{equation}
then, an expansion of Eq.~(\ref{eq28}) with respect to the small
parameters $\delta_{n,s}$ and $k_{0|n,s}/k$ yields in the leading
approximation the following expression for $\delta_{n,s}$ :
\begin{equation}\label{eq32}
    \delta_{n,s}\simeq \frac{j_{n,s}}{Q_0 a} \,
\frac{H_n^{(1)}(Q_0\,a)}{H_n^{(1)\prime}(Q_0\,a)}\; ,
\end{equation}
where $Q_0$ is defined by Eq.~\eqref{les1}.  As one can see, the
imaginary part of $k_{0|n,s}$ (as well as the correction to the real
part) is of order of $\sim (Q_0a)^{-1}$ and is small at least for the
first quasistationary levels provided the condition (\ref{eq29}) is
fulfilled. Thus, for observing a quasistationary level, the wavenumber
inside the obstacle should be equal to the complex eigenvalue
$k_{0|n,s}$ determined by Eq.~(\ref{eq31}). From the relationship
$k_{0|n,s}^2=k_{n,s}^2-Q_0^2$ we find in the leading approximation the
resonance values of the wave vector $k$ of the incident wave:
\begin{equation}\label{eq32a}
k_{n,s}a\equiv (k_{n,s}^{\prime}+{\rm i}\, k_{n,s}^{\prime\prime})a\approx
 Q_0a+\frac{1}{2}\,
\frac{j_{n,s}^2}{Q_0a}+{\rm i}\, \frac{j_{n,s}\delta_{n,s}^{\prime\prime}}{Q_0a},
\end{equation}
where $\delta_{n,s}\equiv \delta_{n,s}^{\prime}+{\rm i}\,
\delta_{n,s}^{\prime\prime}$ and the contribution of the real part
$\delta_{n,s}^{\prime}$ has been neglected with respect to larger
contributions of order $\sim(Q_0a)^{-1}$.

\begin{figure}
\includegraphics*[width=0.99\linewidth]{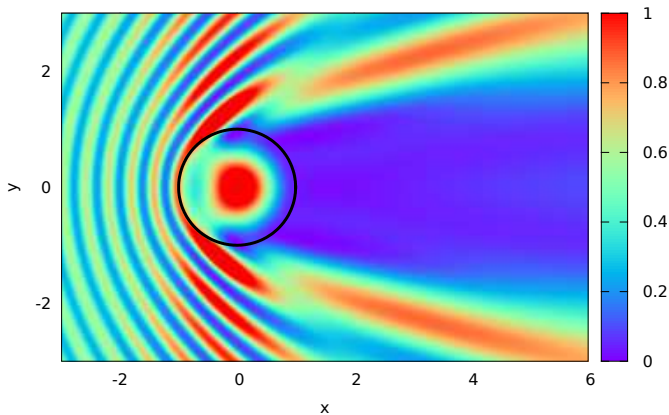}
\caption{Density of the wave function scattered by a
  repulsive disk with $2 m a^2 U_0/\hbar^2 =60$. The incident wave
  vector is $k\,a=8.12$ in order to meet the requirement of
  quasi-resonance in the $s$ channel (cf. the discussion in the
  text).}\label{fig7}
\end{figure}

\begin{figure}
\includegraphics*[width=0.99\linewidth]{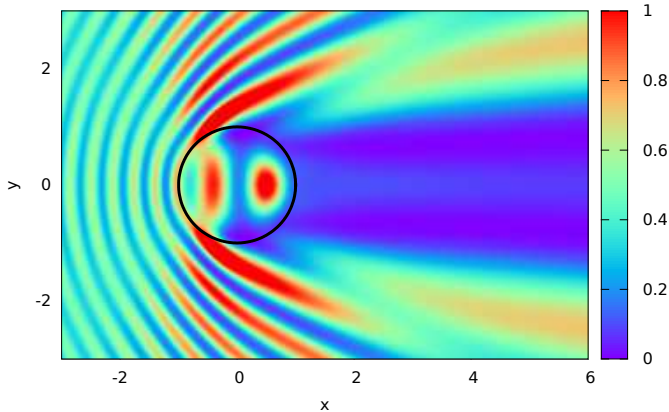}
\caption{Same as Fig. \ref{fig7} for an incident wave vector
$k\,a=8.69$.}\label{fig8}
\end{figure}

As an illustration, we consider a repulsive potential for which the
parameters $U_0$ and $a$ are related by $2 m a^2 U_0/\hbar^2 =60=(Q_0
a)^2$ [thus verifying the condition \eqref{eq29}]. We look for
instance for the first resonance in the $s$-wave channel, i.e., we
consider a configuration where $k_0 a$ is close to the first zero of
$J_0$: $j_{0,1}=2.40482555...$. Formulas \eqref{eq31}, \eqref{eq32}
and \eqref{eq32a} yield $\mbox{Re}\,(k_{0|0,1} a)\simeq 2.385$ and
$\mbox{Re}\,(k_{0,1} a)\simeq 8.12$. The expected resonance is indeed
observed for this value of the incident wave vector, as shown in
Fig.~\ref{fig7}: in this case the density of the scatterer wave has a
strong maximum at the center of the repulsive disk. In Fig. \ref{fig8}
we display the first resonance in the $n=1$ channel. In this case
formula \eqref{eq32a} yields $\mbox{Re}\,(k_{1,1} a)\simeq 8.69$ and
one observes a non-isotropic intensity pattern inside the circle,
typical for a $p$-state.

\begin{figure}
\includegraphics*[width=0.99\linewidth]{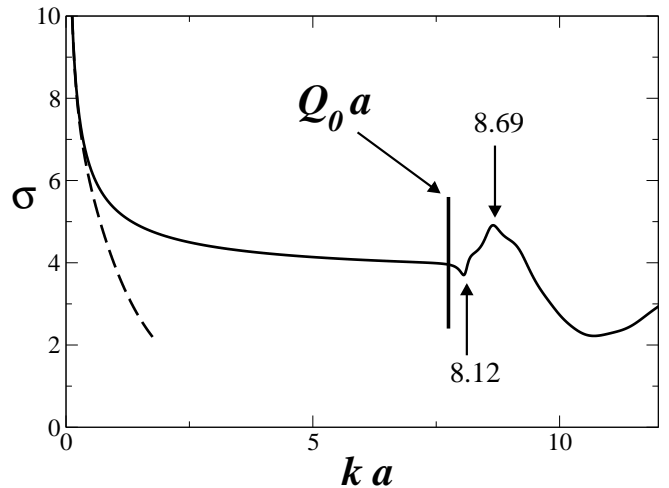}
\caption{Cross section $\sigma(k)$ for a repulsive potential with $2 m
  a^2 U_0/\hbar^2 =60$. The two arrows mark the resonances
  corresponding to the intensity patterns displayed in Fig. \ref{fig7}
  and \ref{fig8}: respectively $k a=\mbox{Re}\,(k_{0,1} a)\simeq 8.12$
  and $k a=\mbox{Re}\,(k_{1,1} a)\simeq 8.69$. The dashed line is the
  low energy approximation described in the text [Eqs. \eqref{les4},
  \eqref{les3} and \eqref{cross2}]. The vertical line locates the
  threshold wavenumber above which $\hbar^2k^2/(2m)>U_0$.}\label{fig9}
\end{figure}

The cross section
\begin{equation}\label{cross1}
\sigma(k)=\frac{1}{4\, k} \sum_{n=-\infty}^{+\infty} |\tilde{B}_n|^2
\end{equation}
for the potential corresponding to Figs.~\ref{fig7} and \ref{fig8}
(for which $2 m a^2 U_0/\hbar^2 =60$) is represented in
Fig.~\ref{fig9}. In this figure the dashed line is the low energy
approximation \eqref{les4} of the cross section, where $\tilde{B}_0$
is given by \eqref{les3} with $\kappa$ being here defined as
\begin{equation}\label{cross2}
\ln\left( \frac{\kappa\,  a \, e^\gamma}{2}\right) =
\frac{1}{Q_0 a} \, \frac{I_0(Q_0 a)}{I_0'(Q_0 a)} \; ,
\end{equation}
which is the version of Eq. \eqref{les2} appropriate for a repulsive
potential ($I_0$ being the modified Bessel function \cite{AS}).  The
two resonances we have identified are marked with arrows in the
figure. One corresponds to a minimum of the cross-section and the
other one to a maximum, following Fano mechanism.

\section{Conclusion}\label{section.conclusion}

In optics one usually devotes a special attention to bright regions
with high intensity of light: focuses, caustics, but also resonances
as in Fig. \ref{fig7} and \ref{fig8}. Another type of singularities
appears in faint light, i.e., close to the dark spots where nodes of
the light field are associated to phase dislocations. In these zones,
complicated phase patterns may occur, with sharp changes (``jumps'')
of phase across certain lines. We believe that such a linear optics
phenomenon was observed in the experiment of Cilibrizzi {\it et al}
\cite{cilibrizzi}.

We have described how the dark elongated valleys observed in
the wake of an attractive obstacle [Ref. \cite{cilibrizzi} and
Fig. \ref{fig2}] can merge and disappear upon changing the incident
wave-vector and/or the depth of the potential. This phenomenon is
accounted for within a fully linear theory and follows a typical
scenario first proposed in Ref. \cite{Nye88}. As a side result,
this shows that the occurrence of a bright region separating two dark
elongated valleys in the wake of the obstacle is not generic. In
particular, contrarily to oblique solitons, it is not observed in the
wake of an impenetrable disk.

The applicability of the approach of the present work is limited by
nonlinear effects caused by the interaction between
polaritons. Considering the interest raised by the observation of
phase defects in optics \cite{optics-singularities}, in Bose-Einstein
condensates \cite{BEC-singulatities}, in polariton condensates
\cite{cilibrizzi,Fla13} and in other fields \cite{Boa06}, it is
appropriate to set up simple criteria making it possible to
discriminate the linear wake from the nonlinear one and to discuss how
nonlinearity affects the structures presented in this article.

\begin{figure}
\includegraphics*[width=0.99\linewidth]{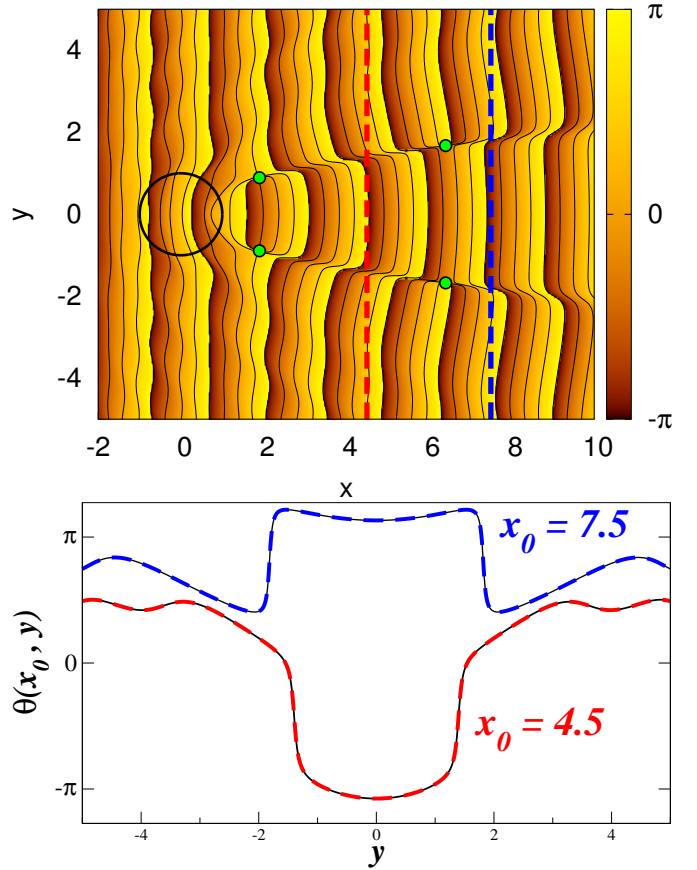}
\vspace*{3mm}
\includegraphics*[width=0.99\linewidth]{fig10b}
\caption{Upper part: color plot of the phase $\theta(x,y)={\rm arg}\,[
  \psi(x,y)]$ for the same potential as in Fig. \ref{fig2} (with also
  $a=1$ and $\vec{k}=4.5\, \vec{e}_x$). The green dots are zeros of the
  wave function. Lower part: detailed plot of the evolution of the
  phase of the wave function along the paths displayed in the upper
  plot as red and blue dashed lines (located at $x_0=4.5$ and
  7.5).}\label{fig10}
\end{figure}

As discussed above, there are marked qualitative differences between
the linear an the non-linear case: the low velocity behavior of the
flows are quite different (superfluid in the nonlinear case and, at
variance, a diverging cross section in the linear case). Also the
comparison between scattering from an impenetrable and a penetrable
defect shows that, whereas the nonlinear wake is expected in both
cases to lead to the formation of oblique solitons, in the linear
regime the dark streaks only appear when the defect is penetrable
(cf. Figs.~\ref{fig1} and \ref{fig2}). Our analysis also provide
another qualitative criterion making it possible to distinguish an
oblique soliton from a dip in a linear wake: the linear dip is
associated to a phase singularity, and the phase of the wave function
varies rapidly when crossing the low density region. This change of
phase corresponds to a minimum or a maximum in the region between the
streaks depending on which side of the singularity one considers. This
is already clear from Fig.~\ref{fig2} and made explicit in
Fig.~\ref{fig10}: along the red dashed line drawn in this figure, the
phase is decreased between two low density regions, whereas along the
blue dashed path the phase is higher when $y\approx 0$. The situation
is quite different for an oblique soliton in the wake of a nonlinear
flow, as can be understood from the following remarks: (i) the phase
of a dark soliton of the Gross-Pitaevskii equation increases in the
direction opposite to its direction of propagation; (ii) an oblique
soliton is stationary because its transverse velocity, related with
the phase jump, is locally compensated by a component of the incident
flow velocity. As a result of properties (i) and (ii) the phase {\it
  always increases} between the low density streaks delimited by two
oblique solitons, as proved theoretically in
Refs.~\cite{egk-2006,Kam08,kk-2011}. Since the phase is measurable in
polariton experiments, this criterion can be implemented in principle
in the analysis of experimental data: nonlinear oblique solitons can
be identified from the fact that the phase always increases between
the low density streaks.

Note also that, in addition to this qualitative discussion, it is very
instructive to perform a quantitative analysis of the change of the
width of the dark strips with the distance from the defect in the
linear and in the nonlinear cases. Studies of this type, with
corresponding experimental data, are presented in the supplementary
material of Refs.~\cite{cilibrizzi} and \cite{amo-2011}.  

We now present quantitative estimates making it possible to evaluate
the importance of nonlinear effects in the structures of the wake
observed behind an obstacle. Working in the standard mean field
approach, interaction effects are most easily accounted for by
replacing the effective Schr\"odinger equation \eqref{eq1} by the
Gross-Pitaevskii equation:
\begin{equation}\label{eq40}
    {\rm i}\,\hbar\,\psi_t=-\frac{\hbar^2}{2m}\nabla^2 \psi
    +\left[U(\vec{r}) + g|\psi|^2\right]\psi\; ,
\end{equation}
where $g$ denotes the effective interaction constant (related to the
$s$-wave scattering length which characterizes the low-energy
inter-particle scattering). The linear effects dominate if the
characteristic size $d$ of the diffraction pattern (say, the width of
the dip around a node) is much less than the healing length $\xi$,
\begin{equation}\label{eq41}
 d\ll\xi,
\end{equation}
where the healing length is defined as
\begin{equation}
 \xi=\frac{\hbar}{\sqrt{2mg\rho_0}}\, ,
\end{equation}
and $\rho_0$ is the characteristic density of the polariton gas.
To obtain a rough estimate of the size $d$, we can use the result of
the Born approximation (\ref{eq8}) which yields a scattering amplitude
of the order $f\sim (Q_0a)^2/\sqrt{k}$ and a characteristic angle
$\varphi\sim (ak)^{-1}$. The phase singularity appears at a distance
\begin{equation}\label{eqx}
x\sim f^2\sim (Q_0a)^4/k
\end{equation}
 and the characteristic width of the dip is
\begin{equation}
d\sim x\,\varphi\sim \frac{(Q_0a)^4}{k^2a} \; .
\end{equation}
Thus, if the parameters characterizing the obstacle and if the
incident wave number are such that this estimate of $d$ satisfies the
condition (\ref{eq41}), then the wave pattern is formed by purely
linear interference effects; otherwise the nonlinear effects prevail
in the formation of the wake structure in the flow.

As an illustration of the experimental relevance of this type of
  quantitative estimate we return to the above discussion where we
  stated that nonlinear oblique solitons can in principle be
  identified from the fact that the phase always increases between the
  low density streaks. For using this criterion in practice one has to
  measure the sign of the phase jump along a long enough segment of
  the experimentally observed dip: precisely along a length greater
  than the estimate $x$ in Eq.~\eqref{eqx} in order to
  make sure that the region around a possible point of the phase
  singularity is not missed in the measurements.

One should also notice that for increasing incident polaritons density, the
linear wave pattern discussed here transforms gradually into the
so-called ``ship wave'' structure \cite{ship-1,ship-2} located {\it
  outside} the Mach cone. On the contrary, the nonlinear oblique
solitons discussed in
Refs.~\cite{egk-2006,Kam08,amo-2011,grosso-2011} are located
{\it inside} the Mach cone. Combining this remark with the above
estimates permits one to distinguish the most important physical effects
which are responsible for the wave pattern observed in the
experiments.

\begin{acknowledgement}
  We thank A. Amo, E. Bogomolny, C. Ciuti, and D. Petrov for useful
  discussions. AMK thanks Laboratoire de Physique Th\'eorique et
  Mod\`eles Statistiques (Universit\'e Paris-Sud, Orsay) where this
  work was completed, for kind hospitality.  This work was supported
  by the French ANR under grant n$^\circ$ ANR-11-IDEX-0003-02
  (Inter-Labex grant QEAGE).
\end{acknowledgement}


\begin{thebibliography}{99}
\bibitem{remark_saddles} The position of the nodes is easily
  determined from the upper plot of Fig. \ref{fig5} which represents
  the density of the wave function and the wavefronts. On the other
  hand, the position of the saddles is more precisely determined from
  the plot of the streamlines (lower part of the figure).
\bibitem{remark6} We recall that in two dimensions an attractive
  potential always has at least one bound state. The case of resonant
  scattering occurs when the energy of the highest bound state is close to
  threshold.
\end{thebibliography}
\end{document}